\documentclass[12pt,preprint]{aastex}

\slugcomment{Not to appear in Nonlearned J., 45.}

\shorttitle{Strong Response Of Very Broad H$\Beta$ in PG\,1416-129}
\shortauthors{Wang \& Li}

\begin{document}


\title{STRONG RESPONSE OF THE VERY BROAD H$\beta$ EMISSION LINE IN LUMINOUS RADIO-QUITE QUASAR PG\,1416-129}


\author{J. Wang\altaffilmark{1}, and Y. Li\altaffilmark{1}}
\email{wj@bao.ac.cn} 
\altaffiltext{1}{National Astronomical Observatories, Chinese Academy of Science}




\begin{abstract}
We report new spectroscopic observations performed in 2010 and 2011
for luminous radio-quite quasar PG\,1416-129. 
Our new spectra with high quality cover both H$\beta$ and H$\alpha$ regions, and show
negligible line profile variation within a timescale of one year.
The two spectra allow us to
study the variability of the Balmer line profile by comparing the 
spectra with the previous ones taken at 10 and 20 years ago. 
By decomposing the broad Balmer emission lines into two Gaussian profiles, 
our spectral analysis suggests a strong response to the continuum level for 
the very broad component, and significant variations in both bulk 
blueshift velocity/FWHM and flux for the broad component. 
The new observations additionally indicate flat Balmer decrements (i.e., too strong H$\beta$ emission) 
at the line wings, which is hard to be reproduced by recent  optically thin models. With these observations we argue that 
a separate inner optically thin emission-line region might not be necessary in the object to
reproduce the observed line profiles.
\end{abstract}





\section{INTRODUCTION}

Because of the limited spatial resolutions of the currently available instruments, the geometry and kinematics of broad-line 
region (BLR) in Active Galactic Nucleus (AGN) is still an important and unresolved problem.
This poor understanding results in an uncertainty for the virial coefficient $f$ that depends not only on
the kinematics, structure and orientation of the BLR, but also on the virial width estimator used. 
$f$ is equal to 3 for a spherical isotropic velocity distribution, while $f$ can be several times larger 
if the BLR can be described as a circular rotation disk. The poor determination of $f$ finally 
causes an uncertainty in the estimation of the
mass of central supermassive black hole (SMBH) of an AGN in the virial assumption 
(e.g., Onken et al. 2004; Kaspi et al. 2005; Marconi et al. 2008; Woo et al. 2010).

Recent spectral profile modelings indicate that the BLR is much more complex than a single virial component.
In addition to the traditional broad component (hereafter BC, $\mathrm{FWHM}\sim10^3\ \mathrm{km\ s^{-1}}$),
a very broad component (hereafter VBC, $\mathrm{FWHM}\sim10^4\ \mathrm{km\ s^{-1}}$) is usually
required to
reproduce the observed line profiles of both H$\beta$ and H$\alpha$ in some AGNs (e.g., Sulentic et al. 2000;
Netzer \& Trakhtenbrot 2007; Hu et al. 2008; Marziani et al. 2009; Zhu et al. 2009; Zamfir et al. 2010).
The physical interpretation of the two components is, however, still under debate at present. Previous studies have frequently
suggested that the VBC likely arises in optically thin clouds that are closer to the central engine than the BC since 1980's (e.g., 
Shields et al. 1995; 
Sulentic et al. 2000; Zamfir et al. 2010). By examining a large sample 
of quasars selected from SDSS, Hu et al. (2008) proposed that the BC (the intermediate-line region in Hu et al. 2008)
can be explained by an inflow because it
is systematically redshifted with respect to the quasar rest-frame.
The comprehensive photoionization models calculated 
by Korista \& Goad (2004) indicate that the VBC can arise from 
inner BLR (i.e., not a distinct emission-line region) where 
the Balmer line emission rate (and also responsivity) is low.    

Sulentic et al. (2000) detected a violent variability in both continuum and H$\beta$ emission 
line in radio-quite luminous quasar PG\,1416-129 (hereafter 2000 spectrum for short). 
By comparing it with the spectrum (hereafter 1990 spectrum for short) taken by Boroson \& Green (1992, BG92),
the 2000 spectrum shows a dramatic disappearance for the BC of the H$\beta$ emission line. 
However, the VBC can be still clearly identified in the 2000 spectrum.
Basing upon these observational facts, Sulentic et al. (2000) argued that a large fraction of the VBC is likely 
produced in optically thin clouds that are closer to the central source.

In this letter, we re-observed the luminous quasar PG\,1416-129 
(R.A.=14$^h$:19$^m$:03$^s$.8, Dec=13\symbol{23}:10$'$:44$''$, J2000, $z=0.129280$) at 10 years after the 
demise of its H$\beta$ BC to examine the variability 
of the H$\beta$ line profile. In addition to H$\beta$, the H$\alpha$ emission line is also covered in our new spectra. 
The observations and data reductions are described in \S2. \S3 presents the results and   
discussion. The $\Lambda$ cold dark matter ($\Lambda$CDM) cosmology
with parameters $h_0=0.7$, $\Omega_{\rm M}=0.3$, and $\Omega_{\Lambda}=0.7$ (Spergel et al. 2007)
is adopted throughout the letter.

\section{SPECTROSCOPIC  OBSERVATIONS AND DATA REDUCTION}

Two new optical spectra were obtained by the 
National Astronomical Observatories, Chinese Academy of Sciences (NAOC) 2.16m telescope
in Xinglong Observatory at January 25, 2010, and March 29, 2011.
The spectroscopic observations were carried out with the 
OMR spectrograph equipped with a back-illuminated SPEC $1340\times400$ CCD as the detector.
The used grating is 600$\mathrm{g\ mm^{-1}}$, and the slit oriented in the south-north 
direction corresponds a width of $2.\symbol{125}0$. This setup finally translates into a spectral resolution of
$\sim4.5\AA$ as measured from the sky emission lines and comparison arcs. This spectral resolution is 
comparable with the previous observations in BG92. 
The blazed wavelength was fixed at 6500\AA\ in both observations in order to
allow the observed spectra cover the wavelength
range from 4600\AA\ to 6900\AA\ in the observer frame (i.e., cover both H$\beta$ and H$\alpha$ 
emission regions). 
The observations were carried out as close to median as possible. 
In each of the observations,
the object was observed successively twice with the exposure time of 
each frame of 3600s. The two frames were combined prior to the extraction to enhance 
the signal-to-noise ratio and to eliminate the contamination of cosmic-rays easily.
The wavelength calibration was carried out by the helium-neon-argon comparison arc taken
between the two successive frames, i.e., at the position being nearly identical 
to that of the object. The observed flux was calibrated by the Kitt Peak National Observatory 
(KPNO) standard stars HD\,117880 and Feige\,56 in the 2010 and 2011 observations, respectively 
(Massey et al. 1988).

The 2-dimensional spectra were reduced by the standard procedures through the IRAF package\footnote{
IRAF is distributed by the National Optical Astronomical Observatories,
which is operated by the Association of Universities for Research in Astronomy, Inc.,
under cooperative agreement with the National Science Foundation.}, 
including bias subtraction, flat-field correction and cosmical-rays removal before the extraction of 
the 1-dimensional spectra. Each extracted spectrum was then calibrated in wavelength and flux by the corresponding 
comparison arc and standard. In the observer frame, the redshifted H$\alpha$ emission profile is strongly 
distorted by the A-band telluric absorption feature at $\lambda\lambda$7600-7630 due to $\mathrm{O_2}$ molecules. 
In order to properly 
model the H$\alpha$ profile, the two telluric features around $\lambda$6800 and $\lambda$7600
were removed from each observed spectrum by the corresponding standard.  
The Galactic extinction was corrected for both spectra by the color excess $E(B-V)$
taken from the NASA/IAPC Extragalactic
Database (NED), assuming the $R_V=3.1$ extinction law of our galaxy (Cardelli et al. 1989).  
The spectra were then transformed to the rest frame, along with the correction of the relativity 
effect on the flux, according to the 
narrow peaks of H$\beta$. Both reduced spectra (hereafter 2010 and 2011 spectra) are 
displayed in Figure 1 by the bottom two curves. The top two curves show the 1990 and 2000 spectra for a comparison.

For each of the four spectra, we modeled the continuum by a broken-powerlaw 
superposed by a broadened \ion{Fe}{2} template using a $\chi^2$ minimization. The template  
provided in BG92 is broadened to have a FWHM that is identical to that of total H$\beta$ line. 
The minimization was carried out 
in the wavelength range free of strong emission lines (e.g., H$\alpha$, H$\beta$,
[\ion{O}{3}]$\lambda\lambda$4959, 5007, \ion{He}{2}$\lambda$4686) of AGNs. The 
continuum fittings are illustrated in Figure 1 by the superposed red lines for the four spectra. 

The isolated emission-lines were then 
profiled by the SPECFIT task (Kriss 1994) in the IRAF package. 
We modeled each emission line by a sum of a set of several Gaussian components.
As an illustration, the modelings in the 2011 spectrum are shown in the bottom-left panel and 
bottom-right panel in Figure 2 for the H$\beta$ and H$\alpha$ regions, respectively. 
The two middle panels show the same as the bottom ones but for the 2010 spectrum.
In addition to the new spectra, the 1990 and 2000 spectra 
were also modeled by the same method in order to examine the variability 
of H$\beta_{\mathrm{BC}}$ and H$\beta_{\mathrm{VBC}}$. 
The deblendings of the H$\beta$ region are illustrated 
in the two top panels in Figure 2.

\section{RESULTS AND DISCUSSION}

The measured emission-line properties and continuum fluxes at 5100\AA\ are tabulated in Table 1. All the fluxes are reported after scaling the spectra to the total [\ion{O}{3}]$\lambda$5007 flux of the 
1990 spectrum, i.e., $F(\mathrm{[OIII]}\lambda5007)=(3.10\pm0.23)\times10^{-14}\ \mathrm{ergs\ cm^{-2}\ s^{-1}}$. All the
uncertainties presented in the table only include the errors resulted from the spectral modelings.

\subsection{Reliability of Spectral Analysis}

The red-shelf of the H$\beta$ line profile of the object is most unlikely contaminated
by the broad \ion{He}{1}$\lambda\lambda4922, 5016$ emission lines (Veron et al. 2002). 
At first, the broad \ion{He}{1}$\lambda5876$ line emission is not detectable in both new spectra.
Secondly, a corresponding red-shelf can be obviously identified in the H$\alpha$ emission line because 
of the clear inflection of the profile. H$\alpha_{\mathrm{VB}}$ 
is strongly redshifted with respect to the narrow core (e.g., Marziani et al. 2009) 
at a velocity shift 
$\Delta\upsilon\approx2100\ \mathrm{km\ s^{-1}}$ in the 2010 spectrum, which is highly consistent with that of H$\beta_{\mathrm{VB}}$,
$\Delta\upsilon\approx2400\ \mathrm{km\ s^{-1}}$. The consistence occurs not only in the velocity shift, 
but also in the line width of $\sim 10^4\ \mathrm{km\ s^{-1}}$. Such consistence can be also identified in
the 2011 spectrum, although the H$\alpha_{\mathrm{VB}}$ bulk velocity is clearly less than that of 
H$\beta_{\mathrm{VB}}$ (Note the large uncertainty for the H$\beta$ line).

We calculate the virial mass of the central SMBH ($M_{\mathrm{BH}}$) 
from the four spectra taken in different epochs since we believe that the blackhole mass
increases negligibly during the 20 years.   
In each epoch, $M_{\mathrm{BH}}$ is estimated 
from H$\alpha$ and H$\beta$ emission lines by 
adopting the calibrations that are based on line width and line luminosity
(e.g., Peterson et al. 2004; Kaspi et al. 2005; Green \& Ho 2005; Bentz et al. 2006; Vestergaard \& Peterson 2006).
The luminosities are obtained after the correction of the local extinction.  
The extinction is inferred from the  
measured narrow-line flux ratio H$\alpha$/H$\beta$ by assuming the standard Case B 
recombination and a Galactic extinction curve with $R_V=3.1$.
The line width measured from the total broad-line profile is usually adopted as a good virial broadening 
parameter because of the produced close virial relationship (e.g., Peterson \& Wandel 2000;  
Onken \& Peterson 2002; Kollatschnny 2003;). By re-analyzing the reverberation-mapping data, Peterson et al. (2004) claimed that 
the best line-width measurement is the line dispersion calculated from the variable part of the spectrum.
In order to measure the FWHMs of the entire broad-line profiles, the
modeled narrow emission-lines are subtracted from each of the observed spectra before the measurements.
In addition to the entire line profiles,
$M_{\mathrm{BH}}$ is also estimated by using the FWHMs from the BCs.   
Note that the $M_{\mathrm{BH}}$ estimation is seriously complicated by
the significant variation in the line width and velocity shift between 2000 and 2010/11 when the BCs are considered only.
The VBCs are usually excluded from the $M_{\mathrm{BH}}$ estimation because of the large relative velocity with respect to
the systematic velocity.

$M_{\mathrm{BH}}$ is calculated from the line widths and luminosities.
The labels V and G are used to denote the calibrations in Green \& Ho (2005) and Vestergaard \& Peterson (2006), respectively.
The calibrations finally yield consistent values of $M_{\mathrm{BH}}$.
For H$\beta$ line, the average estimates are 
($M_{\mathrm{BH}}^{V\beta}$, $M_{\mathrm{BH}}^{G\beta}$)$_{\mathrm{BC+VBC}}$
=($2.5\pm1.3\times10^8$, $1.8\pm0.8\times10^8$)$M_\odot$, and
($M_{\mathrm{BH}}^{V\beta}$, $M_{\mathrm{BH}}^{G\beta}$)$_{\mathrm{BC}}$
=($1.2\pm0.1\times10^8$, $9.2\pm0.9\times10^7$)$M_\odot$, where the formal errors are 
estimated from the observations at the different epochs. 
For H$\alpha$ line,
we have average estimates of ($M_{\mathrm{BH}}^{G\alpha}$)$_{\mathrm{BC+VBC}}$=
$2.8_{-0.3}^{+0.2}\times10^8\ M_\odot$ and
($M_{\mathrm{BH}}^{G\alpha}$)$_{\mathrm{BC}}$=$8.2_{-0.6}^{+0.6}\times10^7\ M_\odot$ from the two new observations.
Note that these values are consistent with each other because an uncertainty of
a factor of 4-5 is produced by the current available scaling relationships 
(Peterson 2008).

\subsection{Variations of BC and VBC}

One can see clearly that
the object shows negligible line profile variation within the time scale of one year according to
our new observations.
The light-travel time cross a spherically symmetric BLR can be evaluated as
$\tau_\mathrm{LT}\approx500L_{\mathrm{H\beta, 42}}^{1/3}f_{-7}^{-1/3}n_{11}^{2/3}$days, where 
$L_{\mathrm{H\beta, 42}}$ is the H$\beta$ line luminosity in unit of $10^{42}\ \mathrm{ergs\ s^{-1}}$, $f_{-7}$ is 
the fill factor of the emission gas in unit of $10^{-7}$, and $n_{11}$ is the number density of the gas in unit
of $10^{11}\mathrm{cm^{-3}}$. The local-extinction corrected H$\beta$ luminosities are 
$L_{\mathrm{H\beta}}\approx1.3\times10^{42}\ \mathrm{ergs\ s^{-1}}$ for both BC and VBC components. 
The luminosities yield a light-travel time $\tau_\mathrm{LT}\sim500$ day for
both emission-line regions, which agrees with the observed fact that the line profile is non-variable within one year.

The nature of the region emitting H$\beta_{\mathrm{VB}}$ is still under debate. 
Many authors believe that the H$\beta_{\mathrm{VB}}$ components detected in many AGNs come from 
a separate inner optically thin emitting region according to the analysis of line ratios and the
line-continuum variation studies (e.g., Ferland et al. 1990; Marziani \& Sulentic 1993; Corbin 1995, 1997; 
Corbin \& Smith 2000; Sulentic et al. 2000; Zhu et al. 2009). 
The spectroscopic monitors previously
claimed that the variations in line cores are stronger than in wings for 
both high-ionization and low-ionization emission lines (e.g., Peterson et al. 1993; 
Kassebaum et al. 1997; Corbin \& Smith 2000), although the opposite cases are found in Ly$\alpha$ 
and \ion{C}{4}$\lambda$1549 lines for high-z quasars (e.g., O'Brien et al. 1989). In an optically
thin cloud, the emissivity of recombination lines is mainly determined by the volume and the density 
of the cloud rather than the incident ionizing continuum, which results in a weak response to the 
continuum level for the line intensity. By contrast, modern photoionization model of AGN's BLR 
suggests that the H$\beta_{\mathrm{VB}}$ component can be attributed to the inner part of 
the traditional BLR, because the calculated responsivity of 
Balmer emission of optically thick clouds increases steeply with the distance from 
the central engine (Korista \& Goad 2004). Snedden \& Gaskell (2007) 
compared their photoionization models with the broad emission-line profiles of both HST and optical
spectra. Their analysis points out that the optically thick model provides better consistence with
the observations for the very broad component, while the optically thin scenario poses many problems. 
Hu et al. (2008) recently proposed another
scenario that the classical BLR produces the H$\beta_{\mathrm{VB}}$ component, meanwhile
the H$\beta_{\mathrm{B}}$ component is produced by an inflow of material.

Comparing our new spectra with the previous ones allows us to identify a strong responsivity to the continuum
for the H$\beta_{\mathrm{VB}}$ component. As the continuum level at 5100\AA\ decreases from 
$5.2\times10^{-16}\ \mathrm{ergs\ cm^{-2}\ s^{-1}\ \AA^{-1}}$ to 
$\sim2\times10^{-16}\ \mathrm{ergs\ cm^{-2}\ s^{-1}\ \AA^{-1}}$, the measured flux of 
H$\beta_{\mathrm{VB}}$ is reduced by a factor of 2-3. The corresponding equivalent widths 
(EWs) maintain a constant value $\sim100$\AA.
All the EWs are measured against the continuum levels at 5100\AA.
As an additional investigation, the upper panel in Figure 3 shows the smoothed line profiles for H$\alpha$ (the red lines) and 
H$\beta$ (the blue lines). The profiles are produced by a subtraction of the modeled narrow-line profiles, 
and are smoothed by a box of 10 pixels.   
We present the average H$\alpha$/H$\beta$ ratios as a function of radial 
velocity in the lower panel. The average values and the corresponding errors are calculated after
binning the spectra over 1000$\mathrm{km\ s^{-1}}$.    
The results from the 2010 and 2011 spectrum are plotted by the red and blue symbols, respectively.
In both observations, the calculated Balmer decrements have a plateau with $\mathrm{H\alpha/H\beta\sim 3}$
from -5000$\mathrm{km\ s^{-1}}$ to +5000$\mathrm{km\ s^{-1}}$, and decrease rapidly in the wings. The models
in Snedden \& Gaskell (2007) indicate that the 
decrements at the wings can not be reproduced by the optically thin photoionization model
in which the predicted Balmer decrement is rather uniformly around the value of Case B recombination.
Snedden \& Gaskell (2007) claimed that the decreased 
line ratios at wings in their sample are best explained by the optically thick model.

The object shows peculiar variation for the H$\beta_{\mathrm{B}}$ component. 
When compared with the 1990 spectrum, the component is decreased by a
factor of 10 in flux in the 2000 spectrum. However, the flux returns to the 1990 level in the 
2010/2011 spectra in which the continuum level is comparable with that of the 2000 spectrum.  
In addition to the flux change, the 2010/2011 spectra show a large variation in the bulk blueshift velocity and line width for
the BC when compared with the previous observations. The velocity and line width in the 2010 (2011) spectrum 
are larger than those of the 1990 (2000) spectrum by nearly an order of magnitude. Putting the variations in flux and
in velocity together suggests a possibility that the H$\beta_{\mathrm{B}}$ line might be emitted from an episodic outflow 
from the central engine. The outflow is at first radially accelerated
at the launch point, which results in the increasing bulk blueshift and line width. The outflow then reaches a terminal velocity
at larger distance where the radiative force and gravitation are negligible, and has a reduced optical depth.     
The dynamical timescale of BLR is 
$\tau_{\mathrm{dyn}}=R_{\mathrm{BLR}}/\Delta V$, where $R_{\mathrm{BLR}}$ is the distance of the BLR from the 
central SMBH and $\Delta V\sim10^3\ \mathrm{km\ s^{-1}}$ is the typical velocity. 
Estimating the distance as $R_{\mathrm{BLR}}=20(L_{5100}/10^{44}\ \mathrm{erg\ s^{-1}})^{0.67}$ ld (Kaspi et al. 2000)
yields a dynamical timescale $\tau_{\mathrm{dyn}}\sim10$years, which does not pose fundamental challenge for the
above scenario. The second possibility is that the object is a ``Double-peaked'' emission-line QSO whose 
broad Balmer line profiles are produced by an accretion disk (e.g., Chen et al. 1989; Gaskell 2010). In contrast with the
object, the typical ``Double-peaked'' emitters show line profiles with two peaks connected by a depressed
plateau at the line center, and show rapid line profile variation with a timescale typical of a few months. Finally, 
the current data can not exclude a possibility that the two Balmer components come from two separated SMBHs
orbiting with their BLRs within a common NLR (e.g., SDSS\,J153636.22+044127.0, Boroson \& Lauer 2009; Lauer \& Boroson 2009). 
Unusual broad line profile (e.g., two displaced peaks or large shifted broad line) is usually used as an indicator
for binary SMBHs candidates (see recent review in Popovic 2011 and the references therein). The binary BLR model
indicates a significant variation of line profile for different BLR orbiting phases. In PG\,1416-129, assuming a 
$M_{\mathrm{BH}}\sim10^8\ M_{\odot}$ for each of the binary SMBHs, the Eq. (2) in Popovic (2011) results in
a distance between the two black holes of $\sim10^{-2}$ pc when the orbital period of $\sim10$years is used. Alternatively,
the significant variability could be produced from the merging of the BLRs when the two black holes are in
the ``sim-detached'' configuration. Long term spectroscopic monitor, especially in X-ray, is required to test which 
hypothesis is correct for the object.

\section{CONCLUSION}

The spectra of luminous radio-quite quasar PG\,1416-129 were re-observed in 2010 and 2011
to study the variation of its broad-line emission. The two spectra show negligible line variation 
within a timescale of one year. 
By re-analyzing the previous spectra, we identify 
a strong response to the continuum level for the very broad component. 
The wide wavelength coverage of the new spectra indicate 
a flat Balmer decrements at the line wings.
With these results, we argue that a 
separate inner optically thin emission line region might not be necessary in the object.

\acknowledgments

We would like to thank the anonymous referee for his/her very useful comments and suggestions.
The authors thank Profs. Todd A. Boroson and Richard F. Green for providing the optical \ion{Fe}{2} template and 
spectrum taken in 1990. We are grateful to Prof. Jack. W. Sulentic for providing the digital spectrum observed in 2000.
Special thanks go to the staff at Xinglong Observatory as a part of National Astronomical Observatories, 
China Academy of Sciences for their instrumental and observational help. The study is supported by the 
Open Project Program of the Key Laboratory of Optical Astronomy, NAOC, CAS.
This search is supported by the NFS of China (grant 10803008), and by the National Basic Research 
Program of China (grant 2009CB824800).

\begin{figure}
\epsscale{.80}
\plotone{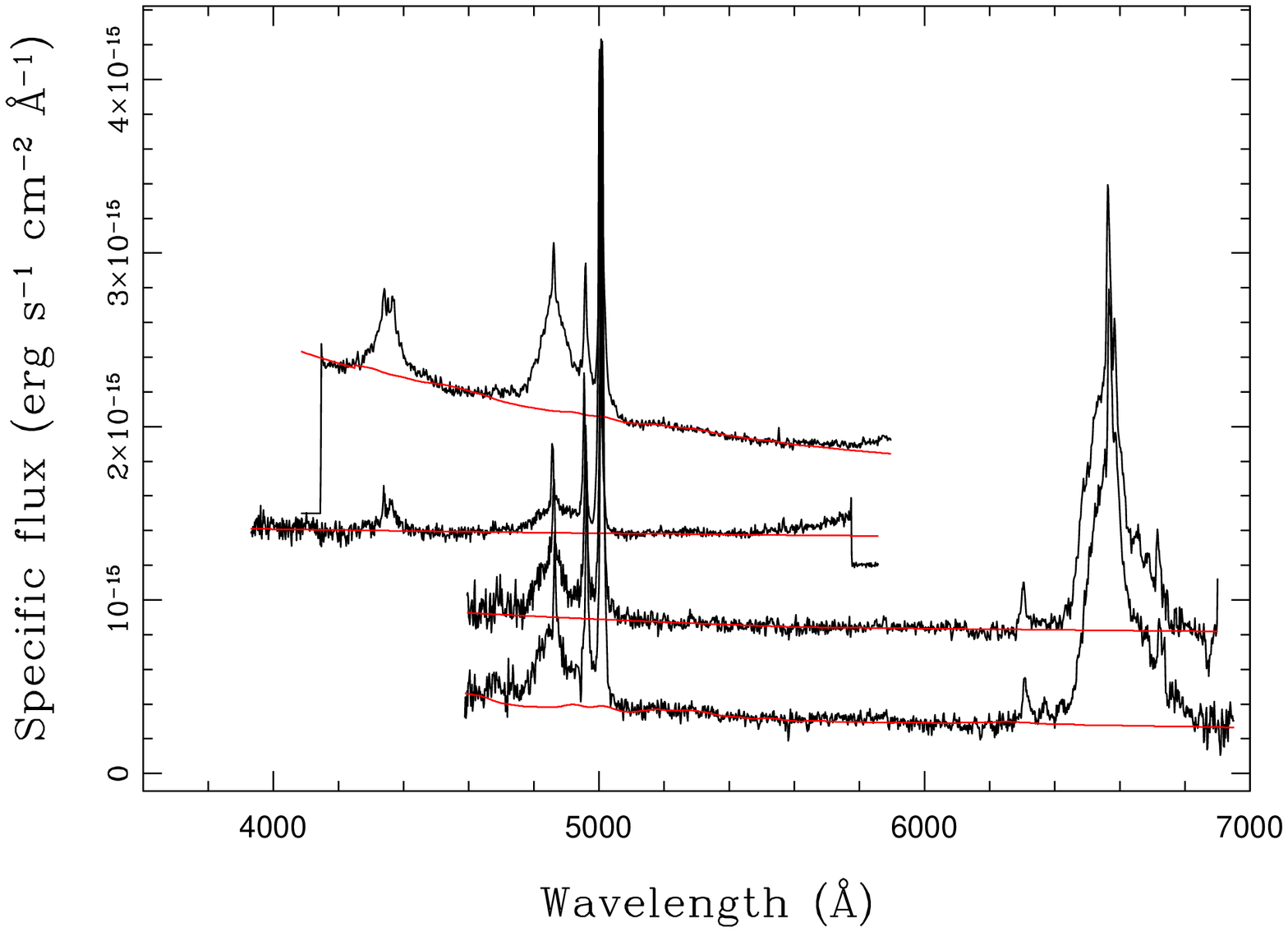}
\caption{The four solid black curves from top to bottom show the observed spectra of PG\,1416-129 taken on 
1990 February 17, 2000 May 08, 2011 May 29, and 2010 January 25. For the latter two spectra,  
both A- and B-band telluric absorption features at $\lambda7600$ and $\lambda6800$ 
are removed from the observed spectrum by the standards. 
The plotted spectra are vertical shifted by an arbitrary amount with respect to the 2010 spectrum for visibility.
The shifted amount is $1.5\times10^{-15}\ \mathrm{erg\ s^{-1}\ cm^{-2}}\ \AA^{-1}$, 
$1.2\times10^{-15}\ \mathrm{erg\ s^{-1}\ cm^{-2}}\ \AA^{-1}$ and 
$5.0\times10^{-16}\ \mathrm{erg\ s^{-1}\ cm^{-2}}\ \AA^{-1}$
for the 1992, 2000, and 2011 spectrum, respectively. 
The modeled continuum and \ion{Fe}{2} complex are 
overplotted by the red curve for each spectrum.  } 
\end{figure}

\begin{figure}
\epsscale{.80}
\plotone{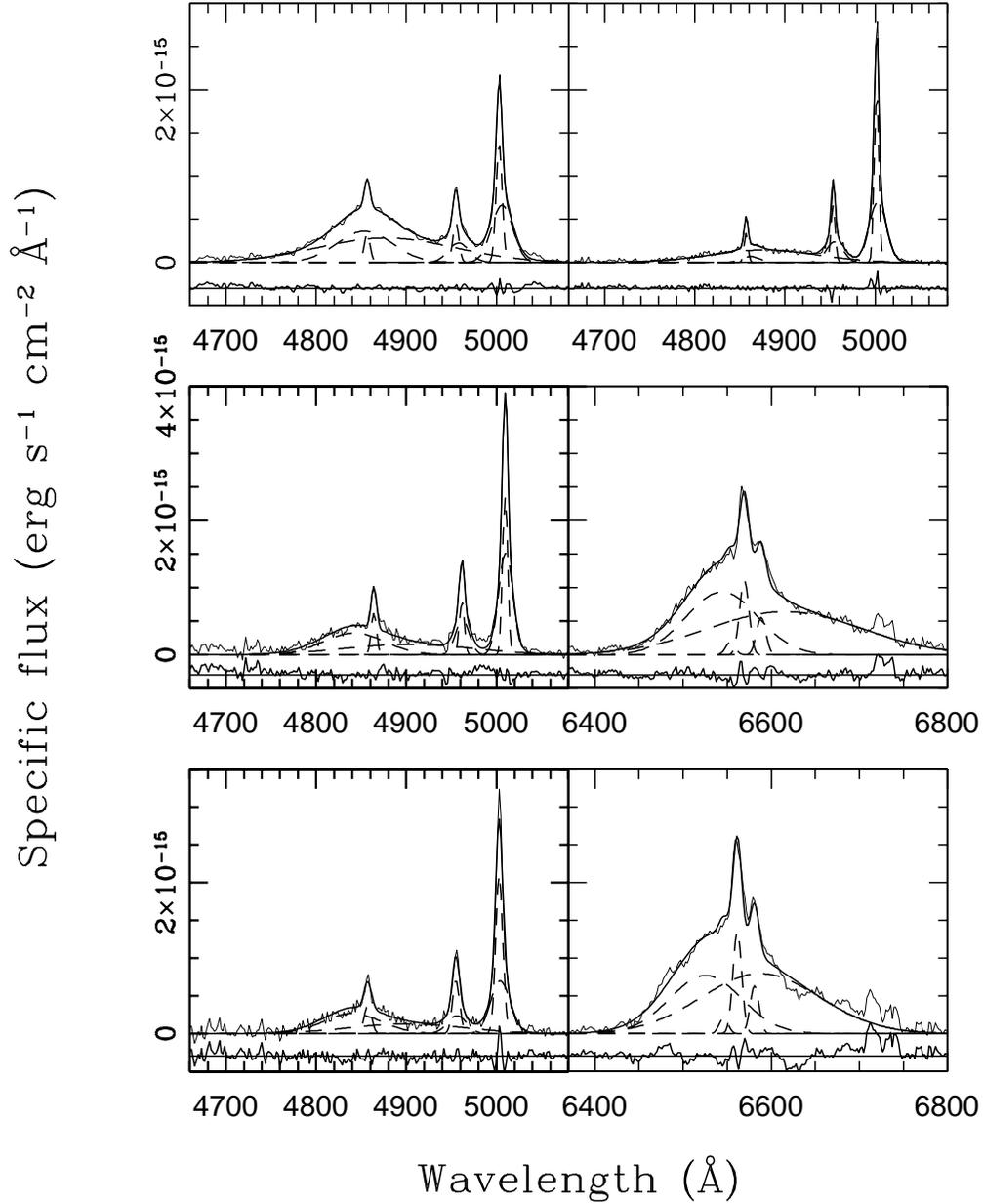}
\caption{Emission-line profile modelings by a sum of a set of Gaussian components for the 1990 spectrum
(top-left panel), the 2000 spectrum (top-right panel),
the 2010 spectrum (middle panels), 
and the 2011 spectrum (bottom panels), after the continuum is subtracted from each of the observed spectra.
For the latter two spectra, the 
left panels show the fitting for the H$\beta$ region, and right panels for the H$\alpha$ region.
In each panel,  
the observed and modeled line profiles are plotted by the light and heavy solid lines, respectively.
Each Gaussian component is shown by a dashed line. The sub-panel underneath each emission-line 
spectrum illustrates the residuals between the observed and modeled profiles.}
\end{figure}

\begin{figure}
\epsscale{.80}
\plotone{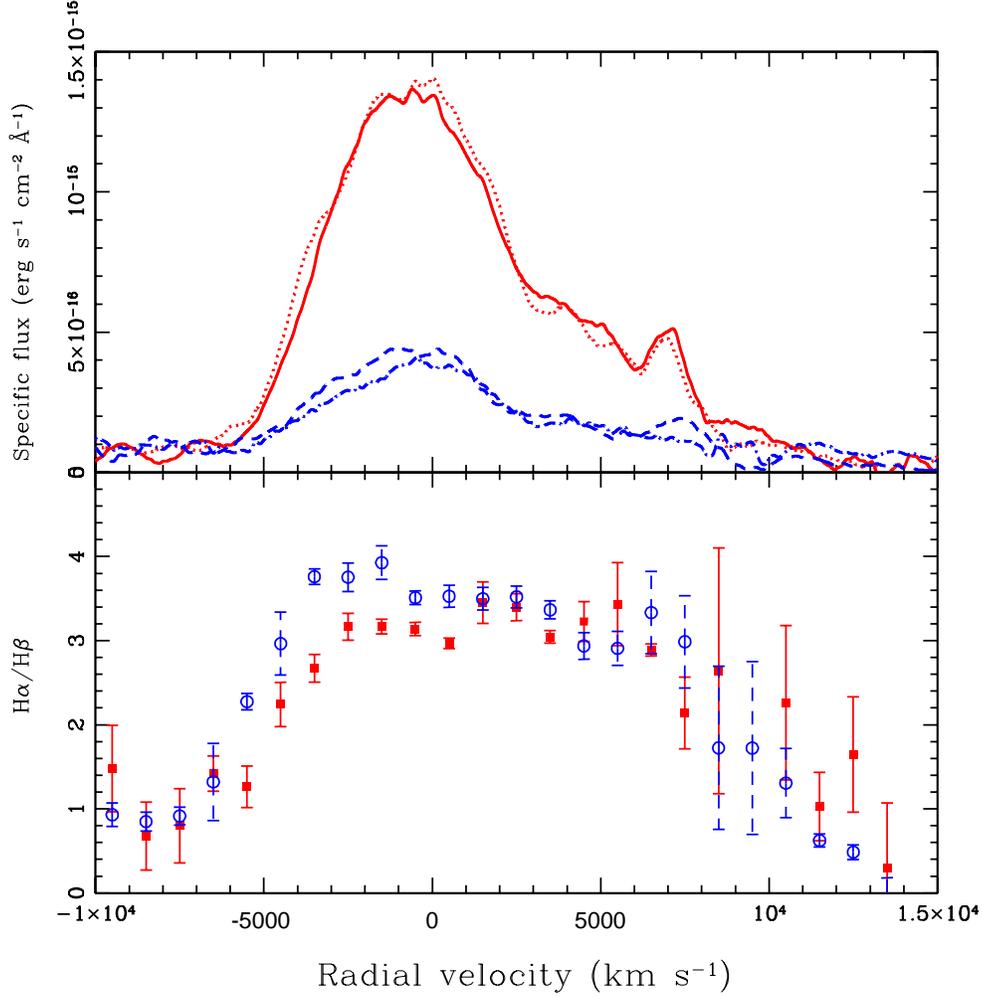}
\caption{\it Upper panel: \rm Smoothed H$\alpha$ (the red lines) and H$\beta$ (the blue lines) line profiles.
The profiles from the 2010 spectrum are shown by the solid and dashed lines. 
The dotted and dot-dashed lines show the profiles from the 2011 spectrum. \it Lower panel: \rm 
Average Balmer decrements across the line profiles for the 2010 (red symbols) and 2011 (blue symbols) spectra.
The values of H$\alpha$/H$\beta$ ratio and the corresponding errors are calculated by binning the spectra over 1000$\mathrm{km\ s^{-1}}$. } 
\end{figure}









\clearpage


\clearpage
\begin{table}
\begin{center}
\caption{EMISSION-LINE MEASUREMENTS FOR PG\,1416-129}
\begin{tabular}{lccc}
\tableline\tableline
 & \multicolumn{3}{c}{1990 FEBRUARY 17\tablenotemark{a}} \\
\cline{2-4} \\
 & Flux & FWHM & $\Delta V$\\ 
LINE IDENTIFICATION  & $10^{-15}\ \mathrm{erg\ s^{-1}\ cm^{-2}}$ & $\mathrm{km\ s^{-1}}$ & $\mathrm{km\ s^{-1}}$ \\
\tableline
H$\beta_\mathrm{B}$\dotfill & $27.8\pm1.4$ & $4483\pm194$ & $-185\pm50$\\
H$\beta_{\mathrm{VB}}$\dotfill & $53.6\pm2.6$ & $10963.6\pm958$ & $1297\pm541$\\
H$\beta_{\mathrm{B+VB}}$\dotfill & $81.4\pm3.0$ & \dotfill & \dotfill\\

\tableline
& \multicolumn{3}{c}{2000 MAY 08\tablenotemark{b}}\\
\cline{2-4} \\
 & Flux & FWHM & $\Delta V$\\ 
LINE IDENTIFICATION  & $10^{-15}\ \mathrm{erg\ s^{-1}\ cm^{-2}}$ & $\mathrm{km\ s^{-1}}$ & $\mathrm{km\ s^{-1}}$ \\
\tableline
H$\beta_\mathrm{B}$\dotfill & $1.9\pm0.8$ & $1443.4\pm441.3$ & $308.0\pm203.$\\
H$\beta_{\mathrm{VB}}$\dotfill & $23.2\pm0.9$ & $8548.1\pm432.7$ & $1244.1\pm227.3$\\
H$\beta_{\mathrm{B+VB}}$\dotfill & $25.1\pm1.2$ & \dotfill & \dotfill\\

\tableline
 & \multicolumn{3}{c}{2010 JANUARY 25 \tablenotemark{c}} \\
\cline{2-4} \\
 & Flux & FWHM & $\Delta V$\\ 
LINE IDENTIFICATION  & $10^{-15}\ \mathrm{erg\ s^{-1}\ cm^{-2}}$ & $\mathrm{km\ s^{-1}}$ & $\mathrm{km\ s^{-1}}$ \\
\tableline
H$\beta_\mathrm{B}$\dotfill &  $18.9\pm1.2$ & $4983\pm248$ & $-1356\pm184$\\
H$\beta_{\mathrm{VB}}$\dotfill & $19.2\pm1.0$ & $10329\pm393$ & $2393\pm191$\\
H$\beta_{\mathrm{B+VB}}$\dotfill & $38.1\pm1.6$ & \dotfill & \dotfill\\
H$\alpha_\mathrm{B}$\dotfill & $63.8\pm4.6$ & $4371\pm166$ & $-1160\pm100$\\
H$\alpha_\mathrm{VB}$\dotfill & $94.2\pm4.1$ & $9301\pm189$ & $2096\pm160$\\
H$\alpha_\mathrm{B+VB}$\dotfill & $158.0\pm6.2$ & \dotfill & \dotfill \\

\tableline
 & \multicolumn{3}{c}{2011 MARCH 29 \tablenotemark{d}} \\
\cline{2-4} \\
 & Flux & FWHM & $\Delta V$\\ 
LINE IDENTIFICATION  & $10^{-15}\ \mathrm{erg\ s^{-1}\ cm^{-2}}$ & $\mathrm{km\ s^{-1}}$ & $\mathrm{km\ s^{-1}}$ \\
\tableline
H$\beta_\mathrm{B}$\dotfill &  $16.6\pm2.2$ & $4614\pm356$ & $-1059\pm192$\\
H$\beta_{\mathrm{VB}}$\dotfill &  $16.0\pm4.5$ & $9117\pm2076$ & $2684\pm486$\\
H$\beta_{\mathrm{B+VB}}$\dotfill &  $32.1\pm1.6$ & \dotfill & \dotfill\\
H$\alpha_\mathrm{B}$\dotfill  & $59.8\pm5.4$ & $4397\pm210$ & $-1642\pm142$\\
H$\alpha_\mathrm{VB}$\dotfill &  $106.0\pm9.4$ & $7183\pm163$ & $1200\pm150$\\
H$\alpha_\mathrm{B+VB}$\dotfill & $165.8\pm10.8$ & \dotfill & \dotfill \\ 

\tableline

\tableline
\tableline
\end{tabular}
\tablenotetext{a}{Specific flux $F_{5100\AA}=5.2\times10^{-16}\ \mathrm{ergs\ cm^{-2}\ s^{-1}\ \AA^{-1}}$.}
\tablenotetext{b}{Specific flux $F_{5100\AA}=2.0\times10^{-16}\ \mathrm{ergs\ cm^{-2}\ s^{-1}\ \AA^{-1}}$.}
\tablenotetext{c}{Specific flux $F_{5100\AA}=2.3\times10^{-16}\ \mathrm{ergs\ cm^{-2}\ s^{-1}\ \AA^{-1}}$.}
\tablenotetext{d}{Specific flux $F_{5100\AA}=3.1\times10^{-16}\ \mathrm{ergs\ cm^{-2}\ s^{-1}\ \AA^{-1}}$.}
\end{center}
\end{table}






\end{document}